\documentclass{emulateapj}

\slugcomment {Accepted to the Astronomical Journal on August 21, 2010}

\begin{document}

\title{Optical Discovery of an Apparent Galactic Supernova Remnant G159.6+7.3}

\author{Robert A.\ Fesen \& Dan Milisavljevic}
\affil{6127 Wilder Lab, Department of Physics \& Astronomy, Dartmouth
  College, Hanover, NH 03755}

\begin{abstract}

Deep H$\alpha$ images of portions of a faint $3\degr \times 4\degr$ H$\alpha$
shell centered at $l = 159.6\degr, b = 7.3\degr$ seen on the Virginia Tech
Spectral Line Survey images revealed the presence of several thin emission
filaments along its eastern limb.  Low-dispersion optical spectra of two of
these filaments covering the wavelength range of 4500 -- 7500 \AA \ show 
narrow H$\alpha$ line emissions with velocities around $-170 \pm 30$ km
s$^{-1}$.  Both the morphology and spectra of these filaments are consistent
with a Balmer dominated shock interpretation and we propose these optical
filaments indicate that the large H$\alpha$ emission shell is a previously
unrecognized supernova remnant.  {\sl ROSAT} All Sky Survey images indicate the
possible presence of extremely faint, diffuse emission from the shell's central
region.  The shell's location more than seven degrees off the Galactic plane in
a region of relatively low interstellar density may account for the lack of any
reported associated nonthermal radio emissions. The rare discovery of a
Galactic SNR at optical wavelengths suggests that additional high latitude SNRs
may have escaped radio and X-ray detection. 

\end{abstract}

\keywords{ISM: supernova remnant - shock waves - X-rays: ISM }

\section{Introduction}

Supernovae (SN) occupy an important place in astrophysics as they are a primary
source of heavy elements in the universe, and generate the creation of 
neutron stars and cosmic rays.  Currently, there are
some 275 Galactic supernovae remnants (SNRs) recognized \citep{Green09} with
additional ones identified or discovered every few years (e.g.,
\citealt{Brogan06}).  

Before the advent of radio astronomy, only two remnants were known via optical
studies; the Crab Nebula associated with SN 1054 and the remnant of Kepler's SN
of 1604 \citep{Minkowski64}.  Today, the vast majority of known Galactic SNRs
were first identified through their nonthermal radio emissions
\citep{Milne70,Downes71}.  With increasingly sensitive observations across a
broader spectrum of wavelengths, particularly in X-rays via satellites, several
recently identified SNRs were discovered through non-radio observations.

The first X-ray discovered Galactic SNR is the radio faint but X-ray bright SNR
G156.2+5.7 which was found through ROSAT all-sky survey
images \citep{Pfeff91} and only later confirmed in the radio \citep{Reich92}.
At present, there are more than a dozen SNRs which were discovered through
their X-ray emission (e.g.,
\citealt{Schwentker94,Seward95,Busser96,Achen98,Bamba03,Yama04}), 
with dozens more suspected candidate X-ray SNRs \citep{Achen99}. 

Although there are many examples where optical observations have helped
identify the SNR nature of suspected radio detected Galactic SNR candidates  (e.g.,
\citealt{Parker04,Stupar07}),  recent instances where a Galactic SNR is first
identified solely through its optical emission are exceedingly rare, with only
one case known to the authors.  That remnant, G65.3+5.7, was found
during a wide-field optical emission-line survey of the Milky Way
\citep{Gull77}. It had been missed in previous Galactic radio SNR surveys due
to a weak and partial shell structure \citep{Reich79}, weak and fragmented
associated H$\alpha$ emission \citep{Sharpless59}, and was only recognized as a SNR
through deep [\ion{O}{3}] $\lambda$5007 optical images.

While hundreds of extragalactic SNRs have been first identified through optical
imaging (e.g., \citealt{MC73,Blair81,MF97,Dopita10}), the rarity of optically discovered
Galactic SNRs is not surprising. Apart from the great sensitivity of Galactic
SNR radio and X-ray surveys,  only 21\% of the 275 known Galactic SNRs have any
associated optical emission \citep{Green09}.  Moreover, while early optical
investigations of SNRs \citep{Rodgers60,vdb73,vdb78} found relatively bright
remnant emissions, recent searches of known remnants typically find only faint
or fragmentary optical features detectable only through deep imaging (e.g.,
\citealt{Zealey79,Mav01,Boumis02a,Boumis02b,Boumis08,SP08,SP09}). 

Despite such long odds, here we report the discovery of an apparent new Galactic
SNR based on its optical emission properties. It appears as a large, faint
shell of emission centered at $l = 159.6\degr, b = 7.3\degr$ in the deep but
low resolution H$\alpha$ emission Virginia Tech Spectral Line Survey (VTSS) of
the Galactic plane \citep{Dennison98,Fink03}. We also show ROSAT All Sky Survey images
which may indicate the presence of very weak, diffuse X-ray emission from the remnant's interior.
Below we present higher resolution H$\alpha$ images of this shell (hereafter
referred to simply as G159) and low dispersion optical spectra and propose this object is
a previously unrecognized Galactic SNR.

\section{Observations}

Narrow passband  (FWHM = 30 \AA) H$\alpha$ images of several regions along
G159's optical limb were obtained in February 2010 with a backside illuminated
2048 x 2048 SITe CCD detector attached to the McGraw-Hill 1.3~m telescope at
the MDM Observatory (MDM) at Kitt Peak.  Typical exposure times 
were 1000 to 2000 s. The CCD's 24 micron size pixels gave
an image scale of $0\farcs508$ and a field of view of approximately $17'$
square. However, to improve the signal-to-noise of the detected filament emissions, we
employed $2 \times 2$ pixel on-chip binning.  Follow-up, low-dispersion optical
spectra of portions of the detected emission filaments were immediately obtained with a
Boller \& Chivens CCD spectrograph (CCDS) on the  Hiltner 2.4~m telescope at
MDM Observatory with a 1.2$\arcsec \times$ 5$\arcmin$ slit and a 150 lines
mm$^{-1}$ 4700 \AA\ blaze grating which yielded  $\sim 10$ \AA\ resolution.

Standard pipeline data reduction of these images and spectra was performed using
IRAF/STSDAS\footnote{IRAF is distributed by the National Optical Astronomy
Observatories, which is operated by the Association of Universities for
Research in Astronomy, Inc.\ (AURA) under cooperative agreement with the
National Science Foundation. The Space Telescope Science Data Analysis System
(STSDAS) is distributed by the Space Telescope Science Institute.}.  This
included debiasing, flat-fielding using twilight sky flats, and dark frame
corrections for the images. 

\section{Results and Discussion}

\begin{figure}
\includegraphics[width=\linewidth]{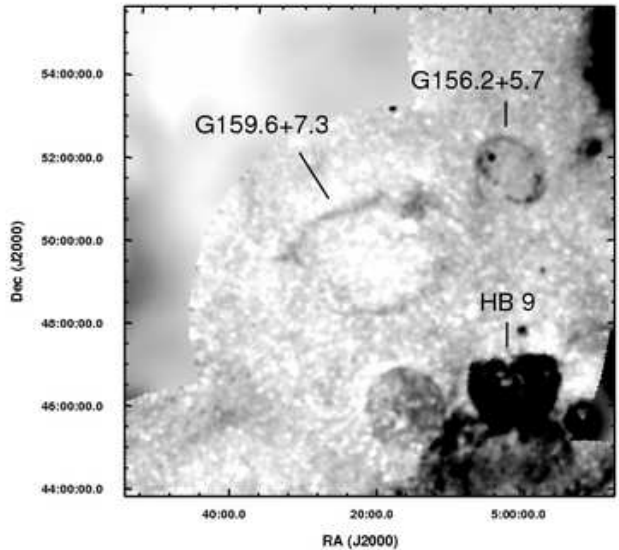}
\caption{H$\alpha$ VTSS image of the emission shell G159.6+7.3. }
\label{fig:figure1}
\end{figure}

Figure 1 shows the G159 emission shell as it appears on the low angular
resolution ($1\farcm6$ per pixel) Virginia Tech Spectral Line Survey (VTSS) of
the Galactic plane. The shell is approximately
$3\degr \times 4\degr$ in angular size and appears nearly complete except for
breaks along its fainter southeastern limb. The shell's H$\alpha$ flux is
extremely weak and is just above the VTSS H$\alpha$ emission measure limit of $2$
cm$^{-6}$ pc along much of its structure. 

Although no filamentary structure is discernible on VTSS image, our higher
resolution H$\alpha$ images of two regions along the shell's eastern limb (see
Fig.\ 2) did reveal several faint filaments. These images (see Figs.\ 3 \& 4)
show an unresolved overlapping filamentary morphology strikingly similar to that
seen in Balmer dominated filaments found in Galactic SNRs where the SN
generated interstellar shocks move through a partially neutral medium
\citep{CR78}. 

\begin{figure}
\includegraphics[width=\linewidth]{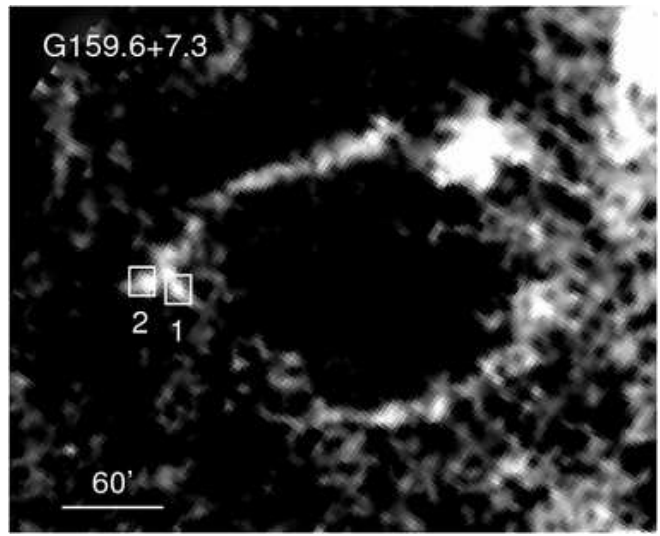}
\caption{H$\alpha$ VTSS image showing the two regions where we obtained higher resolution H$\alpha$ images. 
         North is up, East to the left. }
\label{fig:figure2}
\end{figure}

The long, curved filament shown in Figure 3 is seen to split up into several
thin, fainter and partially overlapping filaments along either end of the
region shown.  At these ends, the filamentary emission fades significantly,
gradually becoming undetectable even in 2000 s long exposures a few arc minutes
away from the region shown.  A second, much shorter but more highly curved
filament is visible in the lower left-hand side of the image. 

The somewhat brighter, N-S aligned filament located in the extreme easterly
region of the shell's limb (see Fig.\ 4) also showed multiple and partially
overlapping thin filaments, only in this case one also finds  considerable
surrounding diffuse emission.  The nature of this diffuse emission is unclear
but its locations near these filaments suggest a possible connection to the
shell.  Despite equally deep H$\alpha$ images taken of several other regions of
the shell, including the relatively bright region along the northwestern limb,
no similar filamentary emission was detected.

\begin{figure*}
\centering
\includegraphics[width=0.6\linewidth]{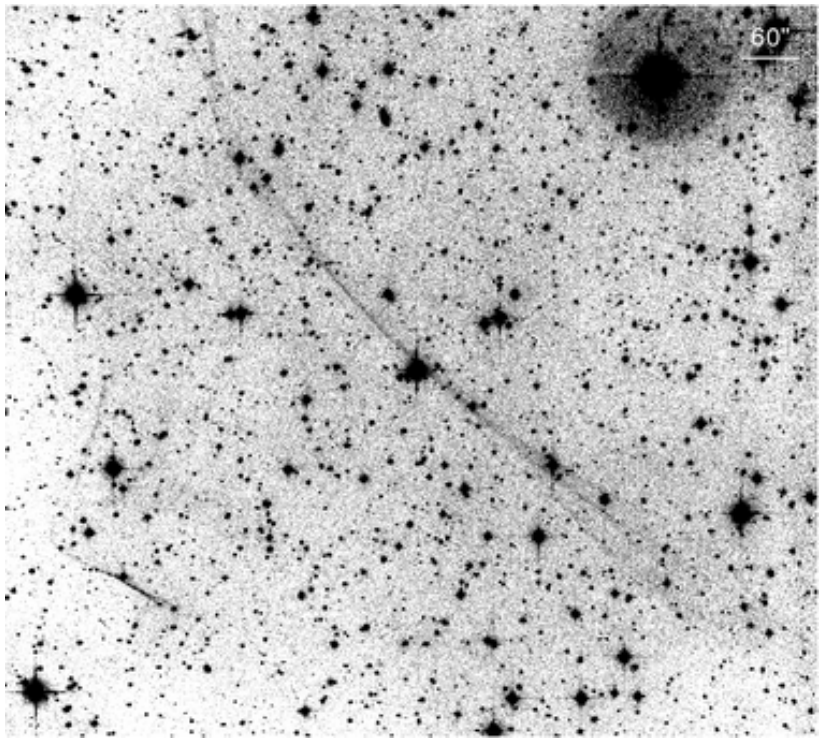}
\caption{H$\alpha$ image of Region 1 in G159.6+7.3. North is up, East to the left. The slit location on the short
southeastern filament where we obtained a low dispersion spectrum (see Fig.\ 5) is marked. Approximate center of image is
$\alpha$(J2000) = $05^{\rm h} 31^{\rm m} 58.0^{\rm s}$,
$\delta$(J2000) = $+49\degr 50' 32''$. }
\label{fig:figure3}
\end{figure*}

\begin{figure*}
\centering
\includegraphics[width=0.6\linewidth]{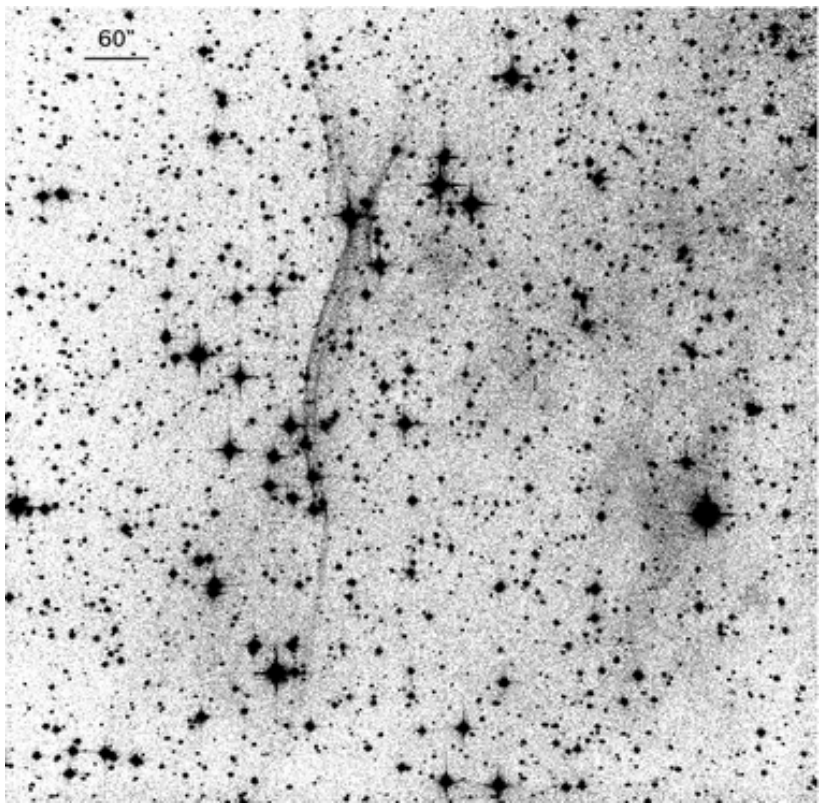}
\caption{H$\alpha$ image of Region 2 in G159.6+7.3.
Approximate center of image is
$\alpha$(J2000) = $05^{\rm h} 34^{\rm m} 12.9^{\rm s}$,
$\delta$(J2000) = $+49\degr 54' 57''$.  North is up, East to the left. }
\label{fig:figure4}
\end{figure*}

Follow-up, low-dispersion optical spectra of the shell's H$\alpha$ filaments
are consistent with a Balmer dominated shock interpretation and thus support an
SNR identification.  In Figure 5, we present our low-dispersion optical
spectrum taken of the bright portion of the short filament seen in the
southeastern corner of Figure 3.  The only emission line seen in its spectrum
is that of narrow (unresolved) H$\alpha$ line emission with no appreciable
[\ion{O}{1}] $\lambda\lambda$6300,6364, [\ion{O}{2}] $\lambda\lambda$7319,7330,
[\ion{O}{3}] $\lambda\lambda$4959,5007, [\ion{N}{2}]$\lambda\lambda$6548,6583,
or [\ion{S}{2}] $\lambda\lambda$6716,6731 emission detected like that commonly
present in photoionized gas (e.g., H II regions and planetary nebulae) or
optical shock filaments \citep{Fesen85}.  The H$\alpha$ emission exhibited
a blueshift of $-170 \pm 30$ km s$^{-1}$ and a flux of $4.6 \times
10^{-16}$ erg s$^{-1}$ cm$^{-2}$. A marginal detection of H$\beta$ can be seen in the spectrum 
suggestive of an H$\alpha$/H$\beta$ ratio $\geqslant$ 4 
implying an $E(B-V) \geqslant 0.35$. 
Such a weak detection of H$\beta$ emission is consistent with a fairly low
amount of extinction of around $E(B-V) \sim 0.45$ toward G159 suggested from H~I
measurements taken in this direction and conversions of N(H) into
$E(B-V)$ values for a typical gas to dust ratio \citep{Bohlin78,Pre95}.

A spectrum was also taken of the roughly N-S aligned filament shown in Figure
4.  Again only a weak detection of unresolved H$\alpha$ emission was seen, with
no detectable [\ion{N}{2}]$\lambda\lambda$6548,6583 or [\ion{S}{2}] emissions
despite integrations of up to one hour.  As before, unresolved H$\alpha$
emission was seen with roughly the same blueshifted velocity as the other
filament, namely $-170 \pm 30$ km s$^{-1}$.

\subsection{A New Galactic Supernova Remnant}

Both the morphology and spectra of the optical filaments seen in Regions 1 and 2 
(Figs.\ 3 \& 4) are consistent with a Balmer
dominated shock interpretation. Consequently, we propose that these optical filaments indicate
that the large G159 emission shell is a previously unrecognized Galactic supernova
remnant.  Considering these filaments' locations along the edge of the large
H$\alpha$ shell, the observed blueshifted velocities of the H$\alpha$ emission
suggest shock velocities in excess of $\sim$200 km s$^{-1}$.  
Although there is no reported nonthermal radio
emission  suggesting the presence of a SNR, the shell's location
of more than seven degrees off the Galactic plane, and hence maybe in a region
of relatively low interstellar density, may account for the lack of reported
associated radio emission.

\begin{figure}
\includegraphics[width=\linewidth]{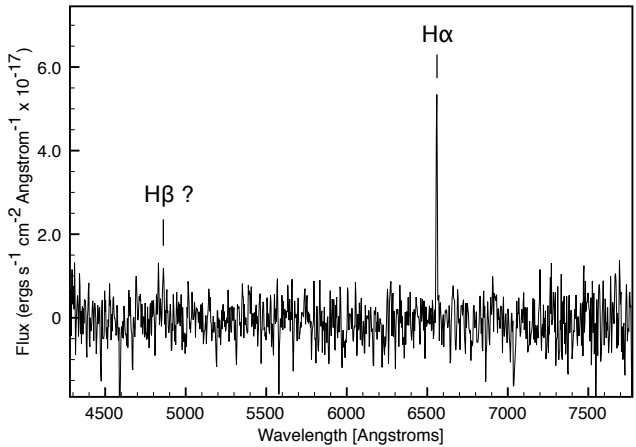}
\caption{Optical spectrum of a relatively bright filament in G159.6+7.3 (see Fig.\ 3) 
         showing narrow H$\alpha$ line emission. }
\label{fig:spectrum}
\end{figure}

With shock velocities above 200 km s$^{-1}$ implied by our optical spectra,
some associated X-ray emission, albeit weak, is anticipated and we have
possibly found faint associated X-ray emission to the G159 shell.  Examination
of on-line {\sl ROSAT} All Sky Survey data (g180p00r5b120pm.fits), shows the possible
presence of extremely faint diffuse emission coincident with the shell's
central region. This is shown in Figure 6 where we present in the upper panel a
wide, low angular resolution {\sl ROSAT} 0.5-2.0 keV image of the G159 region
with several well-known, neighboring Galactic SNRs marked. In the lower panels
of Figure 6, we show smaller regions centered on G159 of the VTSS H$\alpha$
image (left) and the {\sl ROSAT} image (right).

While there are several other similar or even brighter extended emission
patches within the large Galactic plane region shown in {\sl ROSAT} 0.5-2.0 keV
image (Fig. 6, upper panel) which are not know to be associated with any
Galactic SNRs, the rough coincidence of this faint emission patch (barely 
2$\sigma$ above the X-ray background level immediately around it) with G159's
central region is nonetheless interesting.  Such an X-ray morphology would be
consistent with that of a highly evolved SNR where the X-ray emission appears
centrally filled \citep{Shelton99,Williams04}.  Although the X-ray flux from
this emission patch is far too weak with which to produce a meaningful
X-ray spectrum, its positional coincidence, size, and alignment with the G159 shell
lends additional, albeit weak, support to G159's SNR identification.

\subsection{G159's Distance, Size, and Evolutionary Phase}

Although exceedingly faint, this new SNR exhibits several interesting
properties. It would be only the second recently discovered optical SNR and
would support Reich et al.'s (1979) suggestion that many other high latitude SNRs
could have escaped radio detection and identification.  The remnant would also
rank among the very largest Galactic SNRs known in angular size and have
exceptionally weak radio emission.

In the current catalog of the 275 known Galactic SNRs, G159 has no analogue in
the sense that it is not the young remnant of a historic SN (like Tycho or SN
1006) yet shows only Balmer dominated optical emission filaments, a property
mainly seen in young SNRs.  This raises questions about its age: Is it really
old as its large size would suggest with Balmer dominated filaments much like
those seen in parts of the Cygnus Loop, or could it be just a few 1000 yr old and thus
be a larger, older version of the SN 1006 remnant? The SN 1006 remnant is expanding in a very
low density interstellar medium resulting in a high shock velocity  
suppressing the formation of common postshock optical cooling filaments.

\begin{figure*}
\plotone{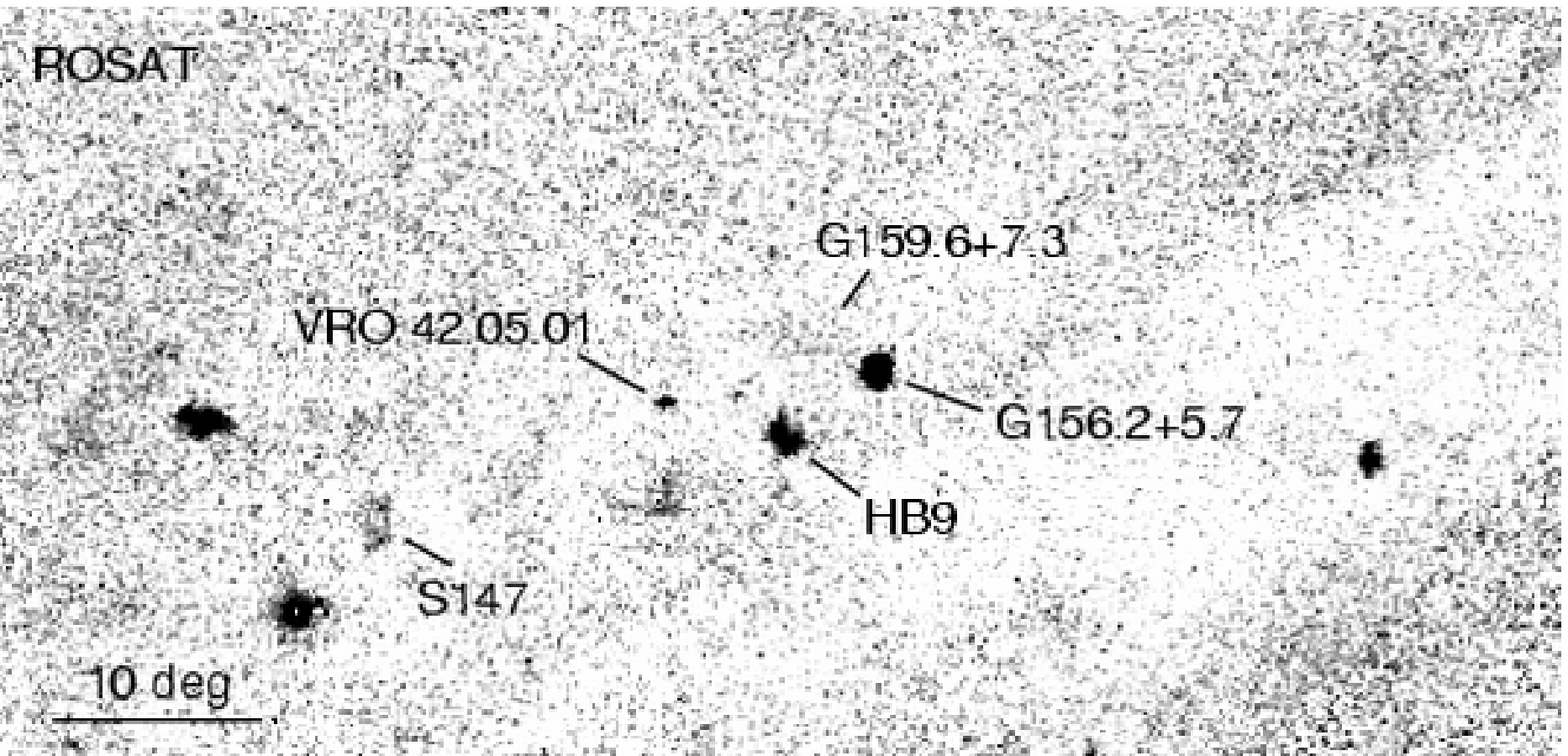}
\plotone{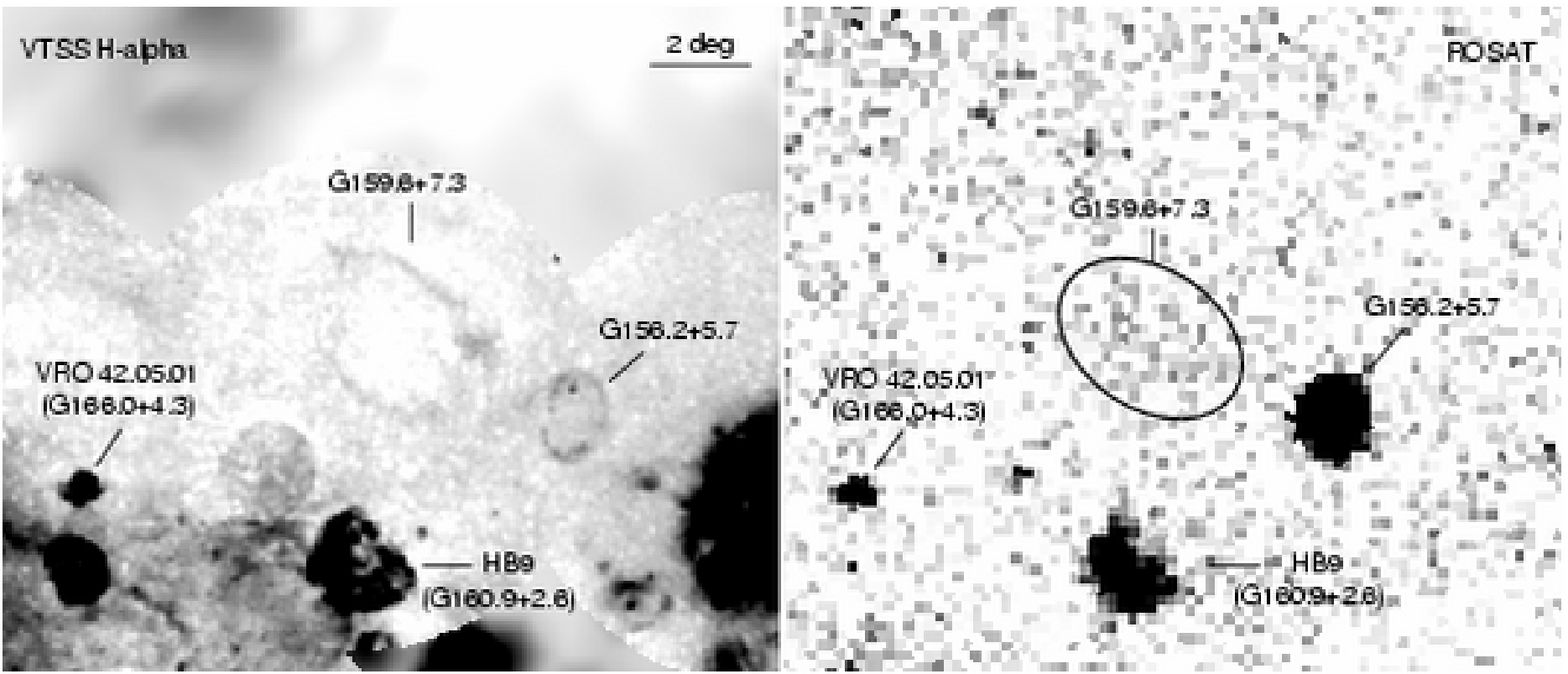}
\caption{The Galactic plane around G159.6+7.3, with ($l,b)$ increasing to the left and top, respectively. 
Top panel shows a wide field view of neighboring SNR X-ray sources as detected
         in the ROSAT All Sky Survey (RASS3). Note the presence of
         faint diffuse emission near the location of G159.6+7.3.  
         Bottom panel shows enlargements of G159.6+7.3 region in the VTSS H$\alpha$ survey (left)
         and ROSAT X-ray image (right). }
\label{fig:x_ray_region}
\end{figure*}

Although G159's distance is unknown, we estimate an upper limit to its distance
of $\simeq$ 2 kpc based on arguments regarding its likely maximum physical
size.  With angular dimensions of $3\degr \times 4\degr$, G159 is about the
same angular size of the Cygnus Loop which lies at an estimated distance of
$\simeq$ 0.54 kpc \citep{Blair05,Blair09}.  If G159 is physically smaller and
hence closer than the Cygnus Loop, then it would rank among the nearest SNRs
known, perhaps just behind the Gum and Vela remnants at $\sim0.25$ kpc.  If, on
the other hand, G159 lies farther away than 1 kpc, its size would substantially
exceed that of the Cygnus Loop and have a diameter in excess of $60$ pc. From 
an observational perspective that there are few remnants with physical diameters
greater than about $100$ pc \citep{Williams04,URO05,Cajko09}, and theoretical
considerations for the radiative expansion phase at diameters around $150$ pc
or less \citep{Cioffi88,SC93,Shelton99}, we suggest G159 lies at a distance
less than 2.5 kpc.  A range of $1 - 2.5$ kpc would place it at a distance much
like the $1.3 - 3.0$ kpc estimated for its brighter neighboring SNR, G156.2+5.7
\citep{Pfeff91,Yama99}.  That remnant is somewhat unusual in that it exhibits,
in addition to several ordinary radiative shock filaments, an extensive set of
Balmer dominated filaments \citep{GF07}, possibly for the same reason as in
G159, namely an expansion into large and relatively low density ISM regions.  

Regardless of its actual distance, G159 is an interesting object.  Being likely
one of the larger Galactic SNRs known, the lack of reported radio emission from
its shell is consistent with SNR evolutionary models which predict a rapid drop
in radio emission in the radiative phase \citep{Asvarov06} and observations of
extragalactic SNRs which suggest weak radio emission for SNRs expanding in low
density media \citep{Pannuti02}.  However, the presence of optical Balmer
dominated filaments raises the question of how such a large and presumably old
SNR has managed to avoid producing any significant optical radiative
emission along virtually its entire large shell structure.  Our failure to
detect any filamentary emission except along the northeastern rim might mean
that its shock velocity is generally less than 70 km s$^{-1}$ and hence too
weak to generate strong optical emission.  However, the true cause of the
paucity of optical filaments, radiative or nonradiative, is currently unknown.
In any case, further study of this newly identified remnant may help resolve its
overall properties which, in turn, may lead to a better estimate of the number
of undetected old, radio and X-ray faint SNRs located well  
off the galactic plane.

\acknowledgements
We wish to thank the MDM staff for their assistance with instrument setups, D.\
Patnaude for helpful conversations, and the referee for suggestions which
improved the paper's presentation.

\end{document}